\documentclass[letterpaper,twocolumn,10pt]{article}

\usepackage{hyperref}
\usepackage{bibunits}
\usepackage{times}
\usepackage{breakurl}
\usepackage[usenames]{color}
\usepackage{tightenum}
\usepackage{multirow}
\usepackage{oneinchmargins}
\usepackage{tightenum}

\usepackage{enumitem}
\usepackage{setspace}
\usepackage{epsfig}
\usepackage{graphicx}
\usepackage{times}
\usepackage{subcaption}
\usepackage{amsmath}
\usepackage[compact]{titlesec}
\usepackage{url}
\usepackage{nonfloat}
\usepackage{multirow}

\usepackage{usenix2019_v3}

  \renewenvironment{thebibliography}[1]{%
    \begin{oldthebibliography}{#1}%
      \setlength{\parskip}{0ex}%
      \setlength{\itemsep}{0ex}%
  }%
  {%
    \end{oldthebibliography}%
  }

\usepackage{color}

\usepackage{listings}
  \usepackage{courier}
\lstdefinestyle{customc}{
  belowcaptionskip=1\baselineskip,
  breaklines=true,
  frame=b,
  xleftmargin=\parindent,
  language=C,
  showstringspaces=false,
  basicstyle=\scriptsize\ttfamily,
  keywordstyle=\bfseries\color{Cerulean},
  commentstyle=\itshape\color{PineGreen},
  identifierstyle=\color{Violet},
  stringstyle=\color{Brown},
}

\lstdefinestyle{customasm}{
  belowcaptionskip=1\baselineskip,
  frame=L,
  xleftmargin=\parindent,
  language=[x86masm]Assembler,
  basicstyle=\footnotesize\ttfamily,
  commentstyle=\itshape\color{purple!40!black},
}

\renewcommand{\em}{\it}

\newcommand{\x}{$\times$}
 
\newcommand{\comment}[1]{}
\newcommand{\ignore}[1]{}

\def\cfigure[#1,#2,#3]{
\begin{figure}
\vspace*{0mm}
\begin{center}

\includegraphics[width=3in]{#1} 
 
\vspace*{-3mm}\caption[]{#2
} \label{#3}
 
\vspace*{-5mm}
\end{center}
\end{figure}}

\def\cfigurefour[#1,#2,#3]{
\begin{figure}
\vspace*{0mm}
\begin{center}

\includegraphics[width=4in]{#1} 
 
\vspace*{-3mm}\caption[]{#2
} \label{#3}
 
\vspace*{-5mm}
\end{center}
\end{figure}}

\def\cfiguretemp[#1,#2,#3]{
\begin{figure}
\vspace*{0mm}
\begin{center}

\includegraphics[width=3.5in]{#1} 
 
\vspace*{-3mm}\caption[]{#2
} \label{#3}
 
\vspace*{-5mm}
\end{center}
\vspace*{-2mm}
\end{figure}}

\def\wfigure[#1,#2,#3]{
\begin{figure*}
\vspace*{0mm}
\begin{center}
 \includegraphics[width=\textwidth]{#1} 
 \vspace*{-3mm}\caption[]{#2
} \label{#3}
 
\end{center}
\end{figure*}}

\def\threefigure[#1,#2,#3,#4,#5]{
\begin{figure*}
\vspace*{0mm}
\begin{center}

\begin{tabular}{ccc}
\includegraphics[width=2in]{#1} & \includegraphics[width=2in]{#2} &  \includegraphics[width=2in]{#3} \\
(a) & (b) & (c) \\
\end{tabular}

\vspace*{-3mm}\caption[]{#4
} \label{#5}

\vspace*{-5mm}
\end{center}
\vspace*{-2mm}
\end{figure*}}

\def\dcfigure[#1,#2,#3,#4,#5,#6]{
{
\begin{figure*}
\begin{center}
\begin{minipage}[c]{\columnwidth}{
\includegraphics[width=\columnwidth]{#1} 
\vspace*{0mm}\caption[]{#2} \label{#3} \
}\end{minipage}\hspace*{\columnsep}\
\begin{minipage}[c]{\columnwidth}{
\includegraphics[width=\columnwidth]{#4} 
\vspace*{0mm}\caption[]{#5}\label{#6} \
}\end{minipage}
\end{center}
\end{figure*}
}
}

\def\tableByTable[#1,#2,#3,#4,#5,#6]{
{
\begin{table*}
\begin{center}
\begin{minipage}[c]{3in}{
\centering
{#1}
\vspace*{0mm}\tabcaption[]{#2}\label{#3} \
}\end{minipage}\hspace*{\columnsep}\
\begin{minipage}[c]{3in}{
\centering
{#4}
\vspace*{0mm}\tabcaption[]{#5}\label{#6} \
}\end{minipage}
\end{center}
\end{table*}
}
}

\def\figureByTable[#1,#2,#3,#4,#5,#6]{
{
\begin{figure*}
\begin{center}
\begin{minipage}[c]{3in}{
\centering
\includegraphics[width=\textwidth]{#1}
\vspace*{0mm}\figcaption[]{#2} \label{#3} \
}\end{minipage}\hspace*{\columnsep}\
\begin{minipage}[c]{3.3in}{
\centering
{#4}
\vspace*{0mm}\tabcaption[]{#5}\label{#6} \
}\end{minipage}
\end{center}
\end{figure*}
}
}

\def\tableByFigure[#1,#2,#3,#4,#5,#6]{
{
\begin{figure*}
\begin{center}
\begin{minipage}[c]{4.3in}{
\centering
{#1}
\vspace*{0mm}\tabcaption[]{#2} \label{#3} \
}\end{minipage}\hspace*{\columnsep}\
\begin{minipage}[c]{2.2in}{
\centering
\includegraphics[width=\textwidth]{#4}
\vspace*{-0.35in}\caption[]{#5}\label{#6} \
}\end{minipage}
\end{center}
\end{figure*}
}
}

\def\doublecfigure[#1,#2,#3,#4]{
{
\begin{figure}
\begin{center}
\begin{minipage}[c]{1.5in}{
\begin{center}
\includegraphics[width=1.5in]{#1}
\end{center}
}\end{minipage}\hspace*{1em}\
\begin{minipage}[c]{1.5in}{
\begin{center}
\includegraphics[width=1.5in]{#2}
\end{center}
}\end{minipage}
\vspace*{0mm}\caption[]{#3} \label{#4} \
\end{center}
\end{figure}
}
}

\def\qcfigure[#1,#2,#3,#4,#5,#6]{
{
\begin{figure*}
\vspace*{0.2in}\
\begin{center}
\begin{minipage}[c]{3in}{
\includegraphics[width=3in]{#1} 
\vspace*{-3mm}
}
\end{minipage}\hspace*{0.5in}\
\begin{minipage}[c]{3in}{
\includegraphics[width=3in]{#2} 
\vspace*{-3mm}
}\end{minipage}

\begin{minipage}[c]{3in}{
\includegraphics[width=3in]{#3} 
\vspace*{-3mm}
}
\end{minipage}\hspace*{0.5in}\
\begin{minipage}[c]{3in}{
\includegraphics[width=3in]{#4} 
\vspace*{-3mm}
}\end{minipage}
\end{center}
\caption[]{#5}\label{#6}
\end{figure*}
}
}

\def\twfigure[#1,#2,#3,#4,#5]{
{
\begin{figure*}
\vspace*{0.2in}\
\begin{center}
\begin{minipage}[c]{6.5in}{
\includegraphics[width=6.5in]{#1} 
\vspace*{-3mm}
}
\end{minipage}

\begin{minipage}[c]{6.5in}{
\includegraphics[width=6.5in]{#2} 
\vspace*{-3mm}
}\end{minipage}

\begin{minipage}[c]{6.5in}{
\includegraphics[width=6.5in]{#3} 
\vspace*{-3mm}
}
\end{minipage}
\end{center}
\caption[]{#4}\label{#5}
\end{figure*}
}
}

\def\dwfigure[#1,#2,#3,#4]{
{
\begin{figure*}
\vspace*{0.2in}\
\begin{center}
\begin{minipage}[c]{6.5in}{
\includegraphics[width=6.5in]{#1} 
\vspace*{-3mm}
}
\end{minipage}

\begin{minipage}[c]{6.5in}{
\includegraphics[width=6.5in]{#2} 
\vspace*{-3mm}
}\end{minipage}

\end{center}
\caption[]{#3}\label{#4}
\end{figure*}
}
}

\def\dssfigure[#1,#2,#3,#4,#5,#6]{
{
\begin{figure*}
\vspace*{0.2in}\
\begin{center}
\begin{minipage}[c]{4in}{
\includegraphics[width=4in]{#1}
\vspace*{-3mm}\caption[]{#2} \label{#3} \
}\end{minipage}\hspace*{0.5in}\
\begin{minipage}[c]{2in}{
\includegraphics[width=2in]{#4}
\vspace*{-3mm}\caption[]{#5}\label{#6} \
}\end{minipage}
\end{center}
\vspace*{-0.4in}\
\end{figure*}
}
}

\def\dsfigure[#1,#2,#3,#4,#5,#6]{
{
\begin{figure*}
\vspace*{0.2in}\
\begin{center}
\begin{minipage}[c]{3in}{
\includegraphics[width=3in]{#1}
\vspace*{-3mm}\caption[]{#2} \label{#3} \
}\end{minipage}\hspace*{0.5in}\
\begin{minipage}[c]{3in}{
\hspace*{0.5in}\
\includegraphics[height=3in]{#4}
\vspace*{-3mm}\caption[]{#5}\label{#6} \
}\end{minipage}
\end{center}
\vspace*{-0.4in}\
\end{figure*}
}
}

\def\dsyfigure[#1,#2,#3,#4,#5,#6]{
{
\begin{figure*}
\vspace*{0.2in}\
\begin{center}
\begin{minipage}[c]{2.5in}{
\includegraphics[height=2.5in]{#1}
\vspace*{-3mm}\caption[]{#2} \label{#3} \
}\end{minipage}\hspace*{0.5in}\
\begin{minipage}[c]{2.5in}{
\includegraphics[height=2.5in]{#4}
\vspace*{-3mm}\caption[]{#5}\label{#6} \
}\end{minipage}
\end{center}
\vspace*{-0.4in}\
\end{figure*}
}
}

\def\dyfigure[#1,#2,#3,#4,#5,#6]{
{
\begin{figure*}
\vspace*{0.2in}\
\begin{center}
\begin{minipage}[c]{3in}{
\includegraphics[height=3in]{#1} 
\vspace*{-3mm}\caption[]{#2} \label{#3} \
}\end{minipage}\hspace*{0.5in}\
\begin{minipage}[c]{3in}{
\includegraphics[height=3in]{#4} 
\vspace*{-3mm}\caption[]{#5}\label{#6} \
}\end{minipage}
\end{center}
\vspace*{-0.4in}\
\end{figure*}
}
}

\def\dyoldfigure[#1,#2,#3,#4,#5,#6]{
{
\begin{figure*}
\vspace*{0.2in}\
\begin{center}
\begin{minipage}[c]{3in}{
\epsfysize=2.0in\
\hspace{0.5in}\
\epsfbox{#1}
\vspace*{-3mm}\caption[]{#2} \label{#3} \
}\end{minipage}\hspace*{0.25in}\
\begin{minipage}[c]{3in}{
\epsfysize=2.0in\
\hspace{0.5in}\
\epsfbox{#4}
\vspace*{-3mm}\caption[]{#5}\label{#6} \
}\end{minipage}
\end{center}
\vspace*{-0.4in}\
\end{figure*}
}
}

\def\cfiguredouble[#1,#2,#3,#4]{
\begin{figure}
\vspace*{0.2in}\
\begin{center}
\begin{minipage}[c]{1.5in}{
\epsfxsize=1.5in\
\epsfbox{#1}
}\end{minipage}\hspace*{0.1in}\
\begin{minipage}[c]{1.5in}{
\epsfxsize=1.5in\
\vspace{0.1in}\epsfbox{#2}
}\end{minipage}\vspace*{-0.10in} \caption[]{#3}\label{#4}
\end{center}
\vspace*{-0.4in}\
\end{figure}
}

\def\wpfigure[#1,#2,#3,#4]{
\begin{figure*}
\vspace*{4mm}
\begin{center}

\includegraphics[width=#4]{#1} 

\vspace*{-3mm}\caption[]{#2
} \label{#3}

\vspace*{-5mm}
\end{center}
\end{figure*}}

\def\wprfigure[#1,#2,#3,#4,#5]{
\begin{figure*}
\vspace*{4mm}
\begin{center}

\includegraphics[width=#4, angle=#5]{#1} 

\vspace*{-3mm}\caption[]{#2
} \label{#3}

\vspace*{-5mm}
\end{center}
\end{figure*}}

\def\DoubleFigureWSlide[#1,#2,#3,#4,#5,#6,#7,#8,#9]{
\begin{figure*}
\vspace*{#9}
\begin{center}
\begin{minipage}{#4}
\includegraphics[width=#4]{#1}
\vspace*{-3mm}\caption{#2
}\label{#3}
\end{minipage}
\hspace{2em}
\begin{minipage}{#8}
\includegraphics[width=#8]{#5}
\vspace*{-3mm}\caption{#6
}\label{#7}
\end{minipage}
\vspace*{-5mm}
\end{center}
\end{figure*}
}

\def\DoubleFigureW[#1,#2,#3,#4,#5,#6,#7,#8]{
\begin{figure*}
\vspace*{0in}
\begin{center}
\begin{minipage}{#4}
\includegraphics[width=#4]{#1}
\vspace*{-3mm}\caption{#2
}\label{#3}
\end{minipage}
\hspace{2em}
\begin{minipage}{#8}
\includegraphics[width=#8]{#5}
\vspace*{-3mm}\caption{#6
}\label{#7}
\end{minipage}
\vspace*{-5mm}
\end{center}
\end{figure*}
}

\def\DoubleFigureWHack[#1,#2,#3,#4,#5,#6,#7,#8]{
\begin{figure*}
\vspace*{0in}
\begin{center}
\begin{minipage}{3in}
\includegraphics[width=#4]{#1}
\vspace*{-3mm}\caption{#2
}\label{#3}
\end{minipage}
\hspace{2em}
\begin{minipage}{3in}
\includegraphics[width=#8]{#5}
\vspace*{-3mm}\caption{#6
}\label{#7}
\end{minipage}
\vspace*{-5mm}
\end{center}
\end{figure*}
}

\def\ddcfigure[#1,#2,#3,#4]{
\begin{figure*}
\vspace*{0.2in}\
\begin{center}
\begin{minipage}[c]{\columnwidth}{
\includegraphics[width=\columnwidth]{#1} 
}\end{minipage}\hspace{0.5in}\
\begin{minipage}[c]{\columnwidth}{
\includegraphics[width=\columnwidth]{#2} 
}\end{minipage} \caption[]{#3}\label{#4}
\end{center}
\end{figure*}
}

\def\ddcfigureSlide[#1,#2,#3,#4,#5]{
\begin{figure*}
\vspace*{#5}\
\begin{center}
\begin{minipage}[c]{3in}{
\includegraphics[height=3in]{#1} 
}\end{minipage}\hspace{0.5in}\
\begin{minipage}[c]{3in}{
\includegraphics[height=3in]{#2} 
}\end{minipage}\vspace*{-0.10in} \caption[]{#3}\label{#4}
\end{center}
\vspace*{-0.4in}\
\end{figure*}
}

\def\cxfigure[#1,#2,#3]{
\begin{figure}
\vspace*{4mm}
\begin{center}
 
\epsfxsize=2.5in\
\epsfbox{#1}\
 
\vspace*{-0.10in}\caption[]{#2
} \label{#3}
 
\vspace*{-5mm}
\end{center}
\vspace*{-2mm}
\end{figure}}

\newcommand{\figWidth}{\columnwidth}
\newcommand{\figSep}{0.05in} 

\newcommand{\beforecaption}{\vspace{-.15cm}\begin{spacing}{0.85}}
\newcommand{\aftercaption}{\vspace{-.45cm}\end{spacing}}
\newcommand{\mycaption}[3]{\beforecaption\caption{\label{#1}{\bf #2} \em\small #3}\aftercaption}

\newcommand{\eg}{\textit{e.g.}}
\newcommand{\ie}{\textit{i.e.}}

\newcommand{\gbs}{~Gbps}

\newcommand{\rnic}{RNIC}
\newif\ifremark
\long\def\remark#1{
\ifremark%
        \begingroup%
        \dimen0=\columnwidth
        \advance\dimen0 by -1in%
        \setbox0=\hbox{\parbox[b]{\dimen0}{\protect\em #1}}
        \dimen1=\ht0\advance\dimen1 by 2pt%
        \dimen2=\dp0\advance\dimen2 by 2pt%
        \vskip 0.25pt%
        \hbox to \columnwidth{%
                \vrule height\dimen1 width 3pt depth\dimen2%
                \hss\copy0\hss%
                \vrule height\dimen1 width 3pt depth\dimen2%
        }%
        \endgroup%
\fi}
\remarktrue

\begin{document}

\pagestyle{empty}




\newcommand{\mm}{mm$^2$}
\newcommand{\figtitle}[1]{\textbf{#1}}
\newcommand{\us}{$\mu$s}
\newcommand{\fixme}[1]{{\color{red}\textbf{#1}}}

\definecolor{pink}{rgb}{1.0,0.47,0.6}
\newcommand{\adrian}[1]{{\color{green}\textbf{#1}}}
\newcommand{\laura}[1]{{\color{pink}\textbf{#1}}}
\newcommand{\shinyeh}[1]{{\color{red}\textbf{#1}}}
\newcommand{\ameen}[1]{{\color{blue}\textbf{#1}}}
\newcommand{\arup}[1]{{\color{yellow}\textbf{#1}}}
\newcommand{\hungwei}[1]{{\color{purple}\textbf{#1}}}

\newcommand{\note}[2]{\fixme{$\ll$ #1 $\gg$ #2}}

\newcommand{\myitem}[1]{\item \textbf{#1}}

\twocolumn[
\begin{@twocolumnfalse}
\begin{center}
{\Large\bf A Double-Edged Sword: Security Threats and Opportunities \\in One-Sided Network Communication}
\end{center}
\smallskip
\centerline{Shin-Yeh Tsai, Yiying Zhang}
\smallskip
\centerline{\em Purdue University}

\bigskip
\bigskip
\end{@twocolumnfalse}
]

\begin{abstract}
One-sided network communication technologies such as RDMA and NVMe-over-Fabrics are 
quickly gaining adoption in production software and in datacenters.
Although appealing for their low CPU utilization and good performance, 
they raise new security concerns that could seriously undermine datacenter software systems 
building on top of them.
At the same time, they offer unique opportunities to help enhance security.
Indeed, one-sided network communication is a {\em double-edged sword} in security.
This paper presents our insights into security implications and opportunities of one-sided communication.
\end{abstract}
\section{Introduction}
\label{sec:introduction}

Traditional network communication in datacenters takes a {\em two-sided} form where 
both the sender and the receiver machines' CPU and software (OS or user-level programs)
are involved in processing network requests.
Several recent technologies in datacenters enable {\em one-sided} 
communication where the receiving machine's CPU and software are completely bypassed
in the processing of incoming network requests. 
Instead, network hardware devices handle incoming requests and directly read/write into 
host machine's memory (\eg, RDMA, Gen-Z~\cite{GenZ-citation}, Omni-Path~\cite{OMNIPATH}), storage (\eg, NVMe-over-Fabrics~\cite{NVMe-fabrics-Inteltalk}), or GPU memory (\eg, GPUDirect~\cite{GPU-Direct}).

Because of its low CPU utilization and low-latency performance, 
one-sided communication has been adopted in distributed in-memory systems~\cite{FARM,Crail},
fast storage systems~\cite{NVMe-fabrics-Inteltalk},
and new architectures like resource disaggregation~\cite{TheMachine,LegoOS}.
Prior research efforts and production systems have focused on
one-sided communication's performance, scalability, programmability, and network protocols,
ignoring one important factor in datacenters, {\em security}.

We call for attention to the security aspects of one-sided network communication.
We study the basic communication patterns
and real hardware implementations of one-sided communication to understand its implications in security.
Based on our findings, we draw several key insights in fundamental issues, challenges, 
and potential solutions in achieving secure one-sided communication.
The most interesting high-level insight is that one-sided communication is a double-edged sword in
security: it can cause security threats and offer opportunities in enhancing security at the same time.

First, one-sided communication's feature of not having any software processing at 
the receiving side poses unique threats in distributed systems.
For example, malicious clients can read or write to remote storage servers with 
one-sided network requests without being noticed, making it hard to account for errors.
Attackers can also easily launch Denial of Service attacks using one-sided communication to swamp
the network or the target machine's memory or storage resources.

Second, in order to bypass host processors, hardware devices that enable one-sided communication
have to implement functionalities that are traditionally built in software.
We identified several security issues in one-sided hardware,
some caused by features added for better performance, some by the need to bypass host CPU,
and some by ill-designed implementation.

Finally, one-sided communication provides a good opportunity to hide accesses from software,
making it easier to achieve user privacy in datacenters.
We designed a secure data store system that leverages one-sided communication to enhance the 
performance of traditional oblivious RAM systems while delivering the same security guarantees.

This paper makes the following contributions:

\begin{itemize}

\item As far as we know, this is the first work to discuss security aspects in one-sided communication.

\item We identified fundamental limitations in one-sided communication that can pose security threats.

\item We discovered new security issues in real one-sided hardware devices. 

\item We leverage one-sided communication to enhance the performance of an ORAM system.

\end{itemize}
\section{One-Sided Communication in Datacenters}
\label{sec:background}

This section provides a background on one-sided communication
and applications built based on it.

\noindent{\bf One-Sided Communication and Enabling Technologies.}
One-sided communication is a type of network communication pattern 
where only the sender side's software is involved in network request processing
but not the receiver side (thus the term ``one-sided'').
One-sided communication can take different forms with different technologies and hardware.
The most famous and popular technology is RDMA, 
a network technology that allows senders to access memory on remote machines directly
without the involvement of remote machines' CPUs.
Similar to RDMA, Omni-Path~\cite{OMNIPATH} is a high-performance fabric architecture 
developed by Intel to support one-sided network communication.
Gen-Z~\cite{GenZ-citation} is another interconnect technology designed for fast, direct access 
to a small (\eg, the scale of a rack) pool of memory devices and it supports one-sided accesses to remote memory. 

Apart from memory-based systems, one-sided communication is also used to access storage systems and accelerators.
NVMe-over-Fabrics~\cite{NVMe-fabrics-Inteltalk} is a technology that 
allows direct network access to NVMe-based storage devices.
GPUDirect~\cite{GPU-Direct} is a technology that allows direct access to remote GPU memory from the network.

\noindent{\bf Applications Based on One-Sided Communication.}
Compared with traditional two-sided communication,
one-sided communication offers key advantages in reducing CPU utilization and in improving performance.
These benefits fit datacenters' performance and cost needs,
making one-sided communication appealing in recent datacenter systems. 
For example, Microsoft Azure and Alibaba have deployed RDMA in large, production scale~\cite{AZURE-RDMA, ALIBABA-RDMA}.
There has also been a host of research work in RDMA and other one-sided technologies in recent years.

Below, we list a representative set of applications that are based on or support one-sided communication.
These include in-memory key-value stores~\cite{KVDIRECT,FARM, FARM15, Mitchell-ATC13, Mitchell-ATC16},
in-memory databases and transactional systems~\cite{NAMDB,DRTM, DRTMH},
graph processing systems~\cite{Shi-OSDI16, GRAM},
consensus implementations~\cite{Wang-SOCC17, DARE},
distributed non-volatile memory systems~\cite{Mojim, Hotpot, Octopus},
remote swap systems~\cite{Gu-NSDI17, RemoteRegion},
NVMe-based storage systems~\cite{Markussen-ICPP18, Guz-SYSTOR17},
and resource disaggregation systems~\cite{HP-TheMachine}.

All these previous systems focus on the performance, scalability, cost, 
programmability, and correctness of one-sided communication.
The rest of this paper will demonstrate that one-sided communication
presents many new security issues and opportunities 
and we call for attention to the security aspect of one-sided communication.
\section{Threats of One-Sided Communication}
\label{sec:design}

While the processing needs at receiver nodes increase CPU utilization in two-sided communication,
the receiver's processing software stack provides the means to implement various security defenses.
In contrast, the lack of receiver involvement in one-sided network communication
poses several security threats in datacenter environments.
This session discusses two fundamental threats of one-sided communication.

\noindent{\bf Threats to accountability.}
In distributed, multi-tenant datacenter environments, many parties can potentially cause 
errors (\eg, node failures, data corruption, stealing information).
It is important to {\em account} for errors when they happen.
Accountability enables the pinpointing of the party responsible for a problem
and allows other parties to be proven innocent~\cite{Popa-ATC11}.
In a threat model where a {\em server} hosts data that multiple {\em client}s can read and write,
an attacker can be one of the clients that desire to write malicious data
or read other clients' data without being noticed.

Under two-sided communication, the receiver handles all incoming network requests
and can identify their senders, providing a means for accountability.
However, it is fundamentally difficult for one-sided communication to be accountable,
because CPU and software are completely bypassed at the receiving side.
Attackers can exploit this vulnerability to perform one-sided network operations 
without being detected.
For example, in RDMA-based in-memory storage systems that support one-sided writes~\cite{LITE,DRTM,NAMDB},
an attacker client can write malicious data to any locations in the store without being detected.
One possible way benign clients can avoid reading unaccountable data 
is to authenticate the writer of the data with encryption keys. 
However, this mechanism needs a fair amount of computation, 
and the performance overhead is especially large 
considering one-sided networks’ high speeds~\cite{CX6, BLUEFIELD}.

\noindent{\bf Threats in denial of service.}
Without any processing at the receiving side, 
one-sided communication also makes it hard to throttle the speed of network requests from any particular sender.
This limitation can be exploited to launch different types of Denial of Service (DoS) attacks~\cite{rfc4297}.
Our threat model here is a cloud environment where an attacker has one-sided network accesses
to a server that provides some service (\eg, an RDMA-based key-value store, GPU acceleration with
GPUDirect) to multiple clients.
An attacker can flood the network to the server with a large number of one-sided network requests
without being detected.

Another potential DoS attack that is different from traditional network DoS attacks is to 
exhaust a server's hardware resources that are used to provide one-sided communication services.
For example, RDMA NICs ({\em RNIC}s), devices that enable one-sided RDMA operations, store metadata
for network connections and memory spaces in on-NIC SRAM so that RNICs can process incoming 
one-sided requests without involving the host machine. 
Attackers can fill the on-NIC SRAM by accessing many different memory spaces, forcing the receiver's 
RNIC to evict victims' metadata.
%
To make it worse, the server cannot tell which client is the attacker, 
because the attacker's one-sided network traffic cannot be detected by the server.

\noindent{\bf Discussion and defenses.}
Attacks that can successfully exploit the above two vulnerabilities rely on the assumption that 
their traffic cannot be sniffed by trusted parties. 
We believe this assumption to be reasonable because most packet 
sniffing tools require the {\em root} privilege to run and 
datacenter administrators usually prohibit packet sniffing.
For example, packet sniffing is prohibited by \rnic\ by default and
requires the RNIC administrator privilege to change the default configuration~\cite{mellanox-manual, RoCESolutions, tcpdump}.

{
\begin{figure*}[th]
\begin{center}
\begin{minipage}{0.9\figWidth}
\begin{center}
\centerline{\includegraphics[width=0.8\figWidth]{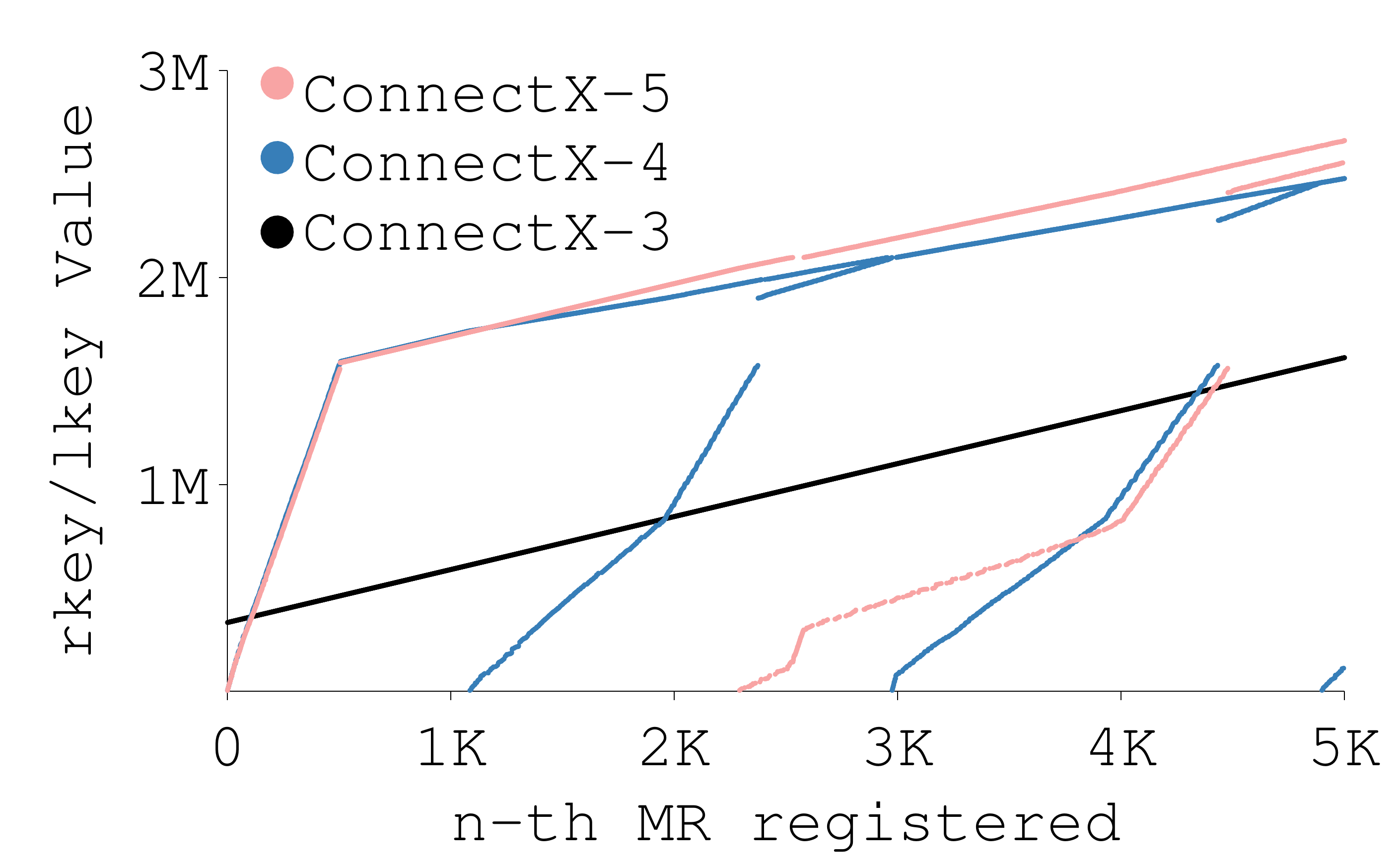}}
\mycaption{fig-rkey}{MR Key Value}
{
X axis represents the sequence of MRs as they are generated in time.
}
\end{center}
\end{minipage}
\begin{minipage}{\columnsep}
\hspace{\figSep}
\end{minipage}
\begin{minipage}{0.9\figWidth}
\begin{center}
\centerline{\includegraphics[width=0.8\figWidth]{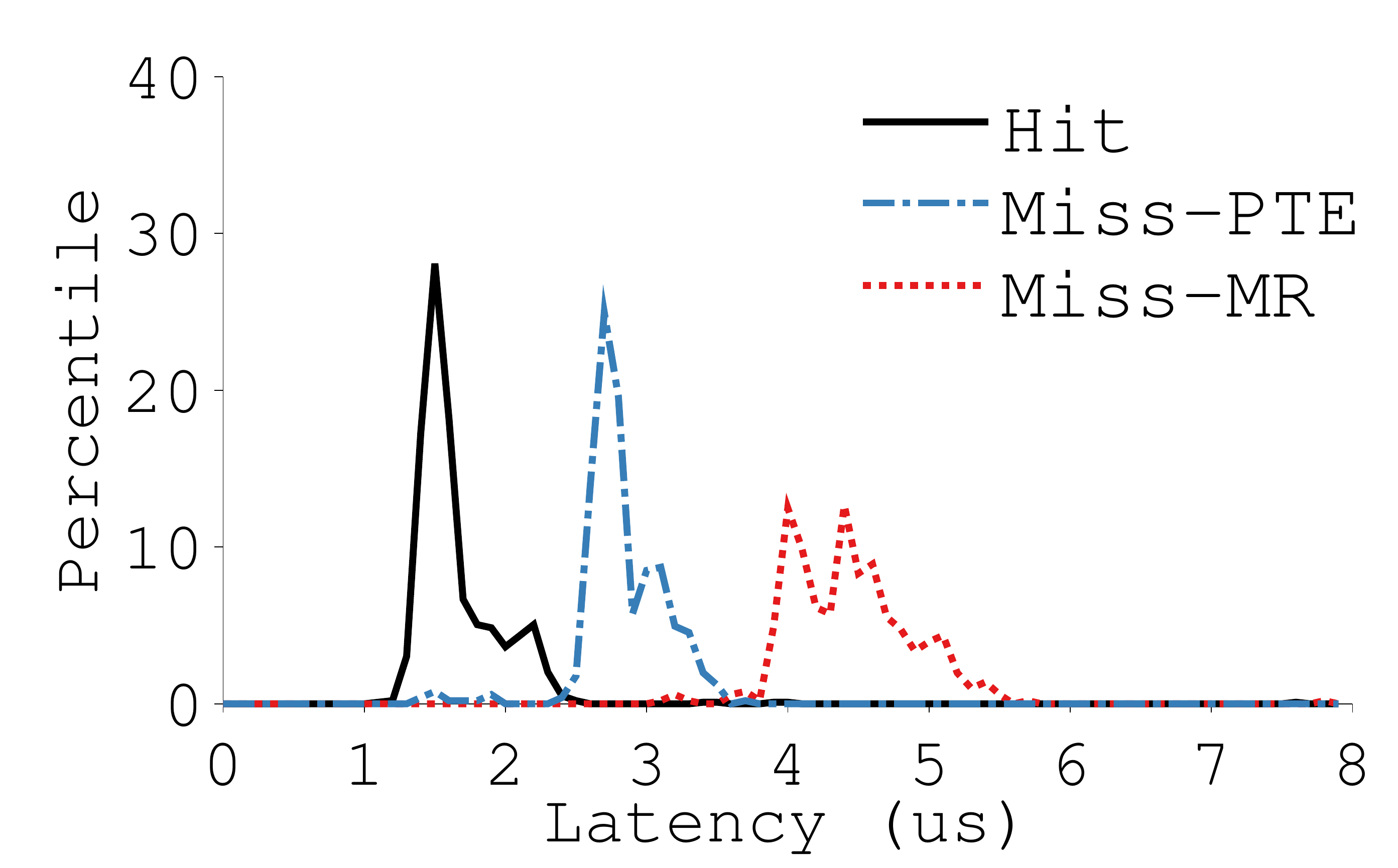}}
\vspace{-0.1in}
\mycaption{fig-timing-diff-cx5}{Timing Differences in ConnectX-5.}
{
}
\end{center}\end{minipage}
\end{center}
\end{figure*}
}

While many traditional defense mechanisms may not work or will break the one-sided communication pattern,
we identified two possible directions for future defenses.
First, at the sender, we can employ defense mechanisms to monitor or control network activities before 
network requests are sent out with the help of an intermediate layer like LITE~\cite{LITE} or with traffic dump tools~\cite{RoCESolutions, mellanox-manual}.
Second, at the receiver, we can enhance the hardware devices that handle one-sided requests with better
security features, for example, by programming SmartNICs.

\section{Vulnerabilities in One-Sided Hardware}
To be able to bypass CPU at the receiver machine, 
hardware devices that enable one-sided communication need to implement many functionalities 
that are traditionally implemented in software such as permission checking and address mapping.
However, hardware is not flexible and most existing one-sided hardware devices are not programmable. 
It is hard to add security features in hardware and security is usually an afterthought in hardware design.

A major threat model we envision is a cloud environment that provides in-memory data store services
to cloud users, similar to the threat model in Section~\ref{sec:design}.
A victim accesses data that a server hosts by issuing one-sided network requests from a client machine.
The attacker runs on another client machine and is another user of the cloud data store service 
(\ie, the attacker has no access to the victim or the server machines).

This section presents three real cases of security issues in one-sided hardware devices.
These security vulnerabilities could potentially be exploited to launch real attacks with the above threat model,
but we leave the design of real attacks for future work.

\noindent{\bf Case 1: hardware-managed ``keys'' are not secret.} 
\rnic{}s use a pair of ``keys'', {\em lkey} and {\em rkey}, 
to protect the local and remote access of each {\em memory region} or {\em MR}.
An \rnic\ of a machine generates and stores lkey and rkey (together with the memory address) at the time
when a user process registers an MR on the machine.
Applications running on other machines use the virtual memory address of a registered MR and its rkey to access it.
When receiving an RDMA request, the RNIC checks its access permission using the rkey.

Surprisingly, we discovered that RNICs generate rkeys in a predictable, sequential pattern with 
Mellanox ConnectX-3, ConnectX-4, and ConnectX-5 RNICs, the three most popular generations of RNICs.
Figure~\ref{fig-rkey} plots the values of lkey/rkey 
of the first 5000 MRs that are registered at a host in the order of the registration time
(lkey and rkey have exactly the same values for all MRs).
For all the three generations of RNICs, there are clear and easy-to-guess patterns in lkey/rkey values.
Moreover, the first 500 MRs have sequentially increasing lkey/rkey values.

Most RDMA-based in-memory storage systems use a small number of large MRs~\cite{LITE,FARM},
making it even easier for attackers to guess MR keys.
After guessing both the rkey and the memory address of an MR (the latter can be guessed in a similar way 
as traditional buffer-overflow attacks),
attackers can gain full access to the MR, overwriting victims' memory content or stealing their data.

\noindent{\bf Case 2: side channels in \rnic{}s.}
A key functionality that all \rnic{}s support for one-sided communication is to map from 
virtual memory addresses to DMA (physical) addresses.
These mappings are essentially page table entries (PTEs) that are stored in host machines' DRAM.
For better performance, \rnic{}s cache hot PTEs in its on-board SRAM.
In addition to PTEs, \rnic{}s also cache MR metadata such as lkeys and rkeys for hot MRs.
An RNIC fetches metadata from its host machine's main memory when receiving RDMA requests whose metadata is not in the RNIC's SRAM.

We evaluated one-sided RDMA requests' latency when both the data's 
PTE and MR metadata is in SRAM (hit), when the data's PTE is not in SRAM (PTE miss),
and when the data's MR metadata is not in SRAM (MR miss).
Figure~\ref{fig-timing-diff-cx5} presents the timing differences of these three cases over
1000 trials of each case.
There is a clear difference between hits and misses, which can be used to establish side channels on \rnic{}s.
Since RDMA accesses from different applications share the same RNIC SRAM,
it is possible to further build real side-channel and covert-channel attacks based on these timing channels.

\noindent{\bf Case 3: exposing physical memory addresses to remote machines.}
Another case where RNIC design can threat security is a feature designed to improve performance.
By default, user processes register MRs with RNICs using virtual memory addresses 
in their address spaces and RNICs store the virtual to physical memory address mapping in their SRAM.
To eliminate the space and performance overhead of address mapping,
Mellanox ConnectX-4 and later versions of Mellanox RNICs introduce a new feature that allows user processes
to register MRs directly using physical memory addresses~\cite{PA-MR}.
Applications running on other machines can use these physical memory addresses to access the MRs.

Exposing physical memory addresses poses many new security threats,
since knowing physical memory address layout is the basis of many attacks.
For example, knowing physical memory addresses make both rowhammer~\cite{Xiao-SEC16, Gruss-DIMVA16, Kim-ISCA14} and throwhammer~\cite{Tatar-ATC18} attacks easier.
Although registering MRs with physical address improves performance, we recommend applications builders to take caution when using this feature.

\noindent{\bf Discussion.}
While the three cases presented in this section all happen with Mellanox RNICs, 
we believe that similar issues can happen with other one-sided hardware and
that there are deep-rooted reasons for why they happen.
There are clear tradeoffs between including more security functionalities in hardware devices and 
the devices' performance and cost. 
For one-sided devices, vendors usually choose the latter over security, because what makes one-sided 
devices appealing is their superior performance and low cost (on saving CPU utilization).
For example, the side channel threats in Case 2 are a result of vendors choosing performance 
(caching hot metadata in on-board SRAM) and cost saving (maximize the utilization of SRAM by not
isolating SRAM space across applications) over security.

Mitigating security issues in one-sided hardware implementation is possible.
For example, \rnic{}s can use cryptographically generated keys as lkey and rkey (Case 1);
they can isolate SRAM for different applications or introduce noise to disturb timing differences 
(Case 2); 
and they can remove the exposure of physical memory addresses (Case 3).
Certain one-sided hardware vendors have manufactured SmartNICs that supports one-sided network
communication~\cite{BLUEFIELD, INNOVA}.
Various defense mechanisms can be implemented on these SmartNICs (\eg, encryption).
Despite the promise of the above mitigations, 
it is still challenging but important to deliver security features with minimal impacts on performance, cost, and hardware complexity.

\section{Opportunity of One-Sided Communication}
\label{sec:opportunity}

Although one-sided communication poses different threats to datacenter security,
it provides one great opportunity to {\em enhance} security.
Users can ensure their privacy by hiding their network activities from receiving servers
using one-sided communication.
This session discusses how we can leverage this opportunity 
to develop secure and fast data storage systems.

\subsection{Environment and Threat Model}
\label{sec:oramthreatmodel}

{\em ORAM}, or {\em oblivious} RAM, is a type of technology that makes access patterns ``oblivious''
by continuously re-encrypting and reshuffling data blocks at storage servers~\cite{Goldreich-87, Goldreich-96, Pinkas-10, Goodrich-12, Shi-ASIA11, Kushilevitz-SODA12, Stefanov-NDSS12, Stefanov-CCS13, Wang-CCS14}.
There have been various secure storage systems built based on ORAM in the past~\cite{Cecchetti-CCS17, Lorch-FAST13, Sahin-SP16,Sahin-SP16, Stefanov-SP13, Williams-CCS12}.
One widely used ORAM system is Path ORAM~\cite{Stefanov-SP13}.
The basic idea of Path ORAM is to organize the server storage
as a binary tree where each node holds a number of (encrypted) objects (\eg, key-value pairs).
Each client caches a small amount of data locally in a {\em stash}.
Client read and write requests are handled in the same Path ORAM protocol to make reads and writes indistinguishable.
To perform a client request (read or write) to an object, the client identifies the path (nodes from a leaf to the root) that contains 
the object using a locally stored mapping table.
The client then reads the entire path into its local stash ({\em read step}).
Afterwards, it remaps the requested object to a random leaf node
and writes all the objects in the path it has read in the read step (with a new value if it is a client write request).
If there are other objects on the path in the client's stash, 
they can also be written back together.

We adopt the same trust model as previous ORAM systems~\cite{Sahin-SP16, Stefanov-SP13, Williams-CCS12, Lorch-FAST13},
where trusted clients access an untrusted storage service provider (\eg, in-memory key-value store, NVMe-based storage).
Clients either directly access the server machine that the service provider runs on through one-sided communication
or access a {\em trusted proxy} which then accesses the service provider through one-sided communication.

We also adopt the same level of ``security'' and ``privacy'' as existing ORAM solutions~\cite{Stefanov-CCS13,Sahin-SP16, Stefanov-SP13, Williams-CCS12, Lorch-FAST13},
where the untrusted service provider should gain no information about client data or access patterns.
Thus, hiding the content of data through encryption alone is not enough.
In addition, no information should be leaked about: 
1) which data is being accessed; 
2) whether the access is a read or a write;
3) when the data was last accessed;
4) whether the same data is being accessed; 
or 5) the access pattern (sequential, random, etc).

\subsection{One-Sided Oblivious RAM}
Although proven to be cryptographically safe,
existing ORAM-based systems have high performance overhead, 
making it too costly to be adopted in many datacenter environments.
We now present our improved ORAM system design. 
Our system is based on Path ORAM~\cite{Stefanov-SP13} but can significantly outperform Path ORAM.

{
\begin{figure}[th]
\begin{minipage}{\columnwidth}
\begin{center}
\centerline{\includegraphics[width=2.6in]{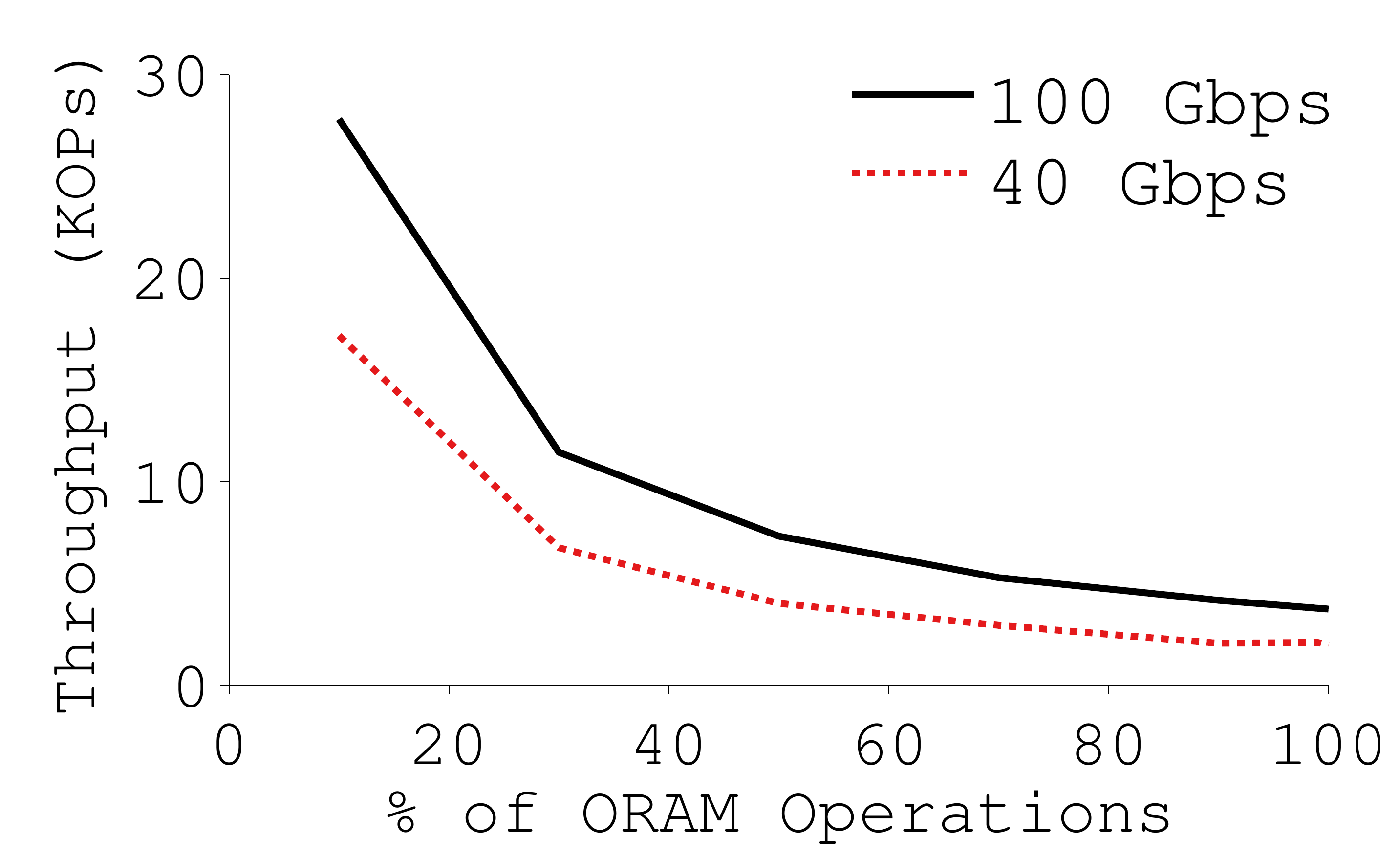}}
\mycaption{fig-oram}{Read Performance of One-Sided ORAM}
{
X axis shows the percentage of read operations that we use original 
ORAM operations to perform.
The two lines show the performance when using a 100\gbs\ and a 40\gbs\ InfiniBand network.
}
\end{center}
\end{minipage}
\end{figure}
}

We propose to leverage one-sided communication to hide data access information and to replace (costly) ORAM 
operations with one-sided operations, thereby improving the performance of ORAM.
Although the basic idea is simple, one needs to take extra care when applying it to provide the same level of security guarantees as ORAM. 
For example, a receiver (the malicious service provider) cannot detect when a one-sided write happens.
But it can take snapshots of data content periodically
and compare two snapshots to detect modifications in between.
It can then infer that a write happens by performing frequent snapshotting. 
Thus, one-sided writes can still leak information to the service provider.
Because of this, we use the unmodified ORAM write protocol and only improve ORAM read performance.

Reading data using one-sided communication can be completely invisible to the service provider, 
even if the provider performs snapshotting. 
For a read-only storage service, we can safely replace all reads with one-sided communication.
However, in practice, most storage services are not read-only
and we need to deliver the second guarantee in Section~\ref{sec:oramthreatmodel} --- 
server should not be able to tell a read from a write.

We use a simple technique to achieve this security goal.
By default, we perform one-sided reads for client read requests
but randomly choose a certain amount of client read requests (\eg, $X\%$ of all read requests)
to perform original ORAM operations.
These ORAM-based read operations cannot be distinguished from write operations.
Therefore, statistically, we can deliver all the security guarantees.
Meanwhile, performing a one-sided read is significantly faster than performing an ORAM operation,
since the latter requires read and write of a whole path, while the former only performs a read to one object.
Although the algorithmic complexity of the modified Path ORAM is still identical to the original Path ORAM, 
the performance of our improved ORAM read is significantly better than the original Path ORAM.

We implemented the original Path ORAM protocol and our modified read mechanism using RDMA
and tested their performance with two network settings, a 40\gbs\ InfiniBand network and a 100\gbs\
InfiniBand network.
Figure~\ref{fig-oram} presents our modified read performance when changing $X\%$ from 10\% to 100\%
(under 100\%, our system falls back to original Path ORAM).
Here, we use a pure-read workload in the YCSB key-value store benchmark~\cite{YCSB, YCSB-C},
with 32,000 512-byte key-value pairs.
We encrypt all data with AES256.
With 50\% one-sided reads, we achieve around 2\x\ performance of pure Path ORAM.

\subsection{Limitations and Discussion}
One limitation of our one-sided ORAM solution is the requirement of prohibiting network packet sniffing. 
As described in Section~\ref{sec:design}, most systems enable this prohibition by default.
However, if an attacker can bypass such prohibition mechanisms (\eg, when it controls a physical machine), 
it can observe one-sided traffic.
Thus, our threat model applies only to other cases, 
for example, when the attacker only owns a virtual machine.

One-sided NICs usually writes to memory through DMA. 
There are methods for a server to track (all) DMA activities 
(\eg, with processor counter monitor~\cite{INTEL-PCM}). 
However, it is still hard to pinpoint one-sided traffic, 
since other types of DMA activities may be happening at the same time. 
Even if a server can detect one-sided network traffic, 
it is still challenging to tell whom the sender is and which data is being accessed.

Our current design does not consider parallel client accesses to the secure data store,
\ie, we only support either a single proxy that delivers all client requests to the data store
or a single client that talks to the data store directly.
Supporting parallel client accesses to a secure data store is an important goal in many distributed
data store systems~\cite{Sahin-SP16, Stefanov-SP13, Williams-CCS12, OBLADI}.
Our design presented here serves as a building block in developing a fully distributed ORAM system.
\section{Concluding Remarks and Future Work}
\label{sec:discussion}

This paper provides the first and initial look into the security aspect of one-sided network communication.
We demonstrate several vulnerabilities that are rooted in one-sided communication's basic design philosophies.
Although we have not built real attacks that exploit these vulnerabilities yet, 
our findings can be a starting point to explore potential attacks and defenses for future researchers and practitioners.

This work is a warning for future one-sided hardware vendors, software developers who want to use one-sided communication,
and datacenters that have or plan to deploy one-sided network systems.
We believe that these three parties should work together to achieve the security goals in datacenters
while preserving one-sided communication's performance and cost benefits.
One promising direction is to leverage programmable network devices like SmartNIC~\cite{BLUEFIELD, INNOVA} and programmable switches~\cite{INCBRICK, NETCACHE, Liang-SIGCOMM18}
to implement security defenses in hardware.

Although adding security guarantees to existing one-sided communication technologies is not an easy job, 
we do not believe the future of one-sided communication to be diminishing.
For all the vulnerabilities that we have identified in this work,
there are defense mechanisms that could potentially work.
Moreover, one-sided communication provides a great opportunity to improve privacy,
the security property that is increasingly important in today's cloud environments.


\vspace{10pt}
{ 
    \begin{spacing}{1.05}
   \bibliographystyle{abbrv}
   \bibliography{all-defs,all,personal,all-confs,local,paper}
   \end{spacing}
}

\end{document}